\documentclass[a4paper,12pt]{article}
\def \bsigma{\mbox{\boldmath $\sigma$}}

\begin{document}
\hspace{11cm}INFN-GEF-TH-4/2000
\vspace {1.5cm}

\begin{center}
\noindent {\bf The general QCD parametrization and the large $N_{c}$ 
description:\\ Some remarks} 

\vskip 30 pt

G.Morpurgo
\vskip 7 pt
Universit\`a di Genova
and Istituto Nazionale di Fisica Nucleare, Genova, Italy.
\end{center}

\vskip 40 pt
\noindent {\bf Abstract.}
Stimulated by a recent paper of Buchmann and Lebed,a comparison is 
presented of the two methods mentioned in the title for treating 
hadron properties in QCD.Doubts arise on the equivalence of the
large $N_{c}$ description to real QCD.
\\
({\em PACS:} 12.38.Aw; 11.15.Pg; 13.40.Dk)

\vskip 40pt 

\baselineskip 24pt
\noindent {\bf 1. Introduction.}
\vskip 5 pt

A recent paper by Buchmann and Lebed (Large $N_{c}$, Constituents quarks 
and $N$, $\Delta$ charge radii) \cite{bl} compares, in a specific 
case, the $1/N_{c}$ description and the general parametrization (GP) 
method of QCD \cite{m,dm,nc}. Ref.[1] seems to imply that, for the 
case at hand, the general parametrization should be looked as a very 
good (though not fully exact) approximation to the $1/N_{c}$ method, 
which is regarded as more fundamental.

I recall that both the $1/N_{c}$ and the GP methods are 
parametrizations to describe hadronic properties, but that the GP 
method, although down to  earth, is founded on QCD, while the same 
is not so clear for $1/N_{c}$. One reason for my doubts on $1/N_{c}$ 
is that expressed concisely in \cite{ss}: ``The basis for the large 
$N_{c}$ approach is the assumption that $N_{c}=3$ QCD is similar to 
QCD in the limit $N_{c}=\infty$. In particular it is assumed that 
there are no phase transitions as we go from $N_{c}=3$ to $N_{c}\to 
\infty$. Currently the status of these assumptions is not clear, 
because not much is known about QCD($N_{c}=\infty$)''.

Note, incidentally, that the factor $g^{2}/3$ producing the hierarchy 
in the $1/N_{c}$ method is, approximately, the same factor that {\em 
empirically} emerges in depressing the diagrams with one added gluon 
in the GP method; so that the two approaches are characterized, in 
practice,by a similar hierarchy.

I will exemplify the general parametrization in a few cases, to 
clarify the situation. But, before doing this, I note two points:

1. The GP method is an exact consequence of QCD, based only on few 
general properties of the QCD Lagrangian. For many physical 
quantities of the lowest multiplet of hadrons (e.g. masses, magnetic 
moments, electromagnetic and semileptonic matrix elements, e.m. form factors 
etc.) it leads to an exact spin-flavor parametrization, independent of 
the choice of the renormalization point of the quark masses in the QCD 
Lagrangian. It turns out that, for a given quantity,  the number of 
terms in this exact QCD parametrization is rather small, indeed 
smaller than one might have anticipated. The GP method -which, even if 
not covariant, is fully relativistic- was developed originally to 
explain the unexpected semiquantitative success of the non 
relativistic quark model (NRQM) \cite{m1}; it did this [2a] long 
before the  $1/N_{c}$ treatment, and much more directly. It 
emerged that the structure of the terms in the GP is similar to that 
of the NRQM. Because terms of increasing complexity in the GP have 
decreasing coefficients, few terms usually  suffice to reproduce the 
data reasonably well, explaining why the NRQM works already in its 
most naive form.

2. Although $SU_{6}$ was important in suggesting the NRQM \cite{m1}, 
it does not play a role after that. For baryons the essential point in 
the construction of the NRQM was that the space part of the octet and 
decuplet wave function has an overall zero orbital angular momentum: 
$L=0$. This implies \cite{m1} the factorizability of the baryon 
(octet or decuplet) NRQM model state as:
\begin{equation}
\phi_B=X_{L=0}({\bf r_1},{\bf r_2},{\bf r_3})\cdot W_B(s,f)
\label{A3}
\end{equation}
where $X$ is the space part and $ W_B(s,f)$ is the spin-flavor factor.
(Color is understood.) The  $W_B$'s are symmetric in the three quark 
variables and have necessarily $J=1/2$ and $J=3/2$ for the octet and 
decuplet, so that, {\em automatically}, the $ W_B$ spin-flavor part 
of $\phi_{B}$ has the form prescribed by $SU_{6}$, without the need of 
invoking $SU_{6}$ at all. The factorizability of $ \phi_{B}$ (1) into 
a space and spin-flavor factor is essential to derive the simple 
structure of the general parametrization. In the GP there is no need 
to relate the states to $SU_{6}$ representations as in the $1/N_{c}$ 
method; nor to rename [1] constituent quarks as ``representation 
quarks''.

Although I will not derive here the GP method -this was done 
repeatedly [2a,3a,4]- I recall some notation, in order to compare GP 
and $1/N_{c}$ in a few cases. The symbol $\vert \phi_{B}\rangle$
indicates, in the quark-gluon Fock space, the state 
corresponding to no gluons and three quarks with wave function 
$\phi_{B}$. The exact eigenstate of the QCD hamiltonian $H_{QCD}$ for 
the baryon $B$ (with mass $M_{B}$) at rest is written  $\vert\psi_{B}\rangle$.
 It is $H_{QCD}\vert \psi_{B}\rangle=M_{B}\vert \psi_{B}\rangle$. 
A unitary transformation $V$ defined in [2a], acting on the auxiliary 
state $\vert \phi_{B}\rangle$, transforms it into the exact 
eigenstate $\vert \psi_{B}\rangle$ of $H_{B}$, so that:
\begin{equation}
	\vert \psi _B\rangle = \vert qqq\rangle + \vert qqq\bar qq\rangle + 
\vert qqq, Gluons\rangle + \cdots
	\label{2}
\end{equation}
where the last form of (\ref{2}) recalls that $V\vert \phi_{B}\rangle$ 
is a superposition of all possible 
quark-antiquark-gluon states with the correct quantum numbers. In 
particular, configuration mixing is automatically included in  
$V\vert \phi_{B}\rangle$. The mass of a baryon is:
\begin{eqnarray}
\label{3}
M_B=\langle \psi_B\vert H_{QCD}\vert \psi_B\rangle =
\langle \phi_B\vert V^{\dag} H_{QCD}V\vert \phi_B\rangle =\nonumber\\
=\langle W_B \vert ``parametrized \  mass"\vert W_B\rangle
\end{eqnarray}
The last step (eliminating the space variables) is due to the 
factorizability of $\phi_{B}$ (eq.(1)). In the next section I discuss 
the ``parametrized mass'' in (\ref{3}).
\vskip 40 pt
\noindent {\bf 2. The parametrization of the baryon masses in the GP 
method.}
\vskip 5 pt

The ``parametrized mass'' in (\ref{3}) following from the GP method is 
[2e,3a]:
\begin{eqnarray}
\label{4}
``parametrized \ mass''=M_{0}\ +\ B\sum_i P_i^s \ +\ C\sum_{i>k} 
(\bsigma_{i} \cdot \bsigma_{k}) \ + \nonumber \\
+\ D\sum_{i>k}(\bsigma_{i} \cdot \bsigma_{k})(P_i^s +P_k^s)\ +\ 
E\!\!\!\sum_{{\scriptsize \begin{array}{c}i\neq k\neq 
j\\(i>k)\end{array}}}\!\! (\bsigma_{i} \cdot \bsigma_{k})P_j^s\ +\ 
a\sum_{i>k}P_i^s P_k^s\ + \\
+\ b\sum_{i>k}(\bsigma_{i} \cdot \bsigma_{k})P_i^s P_k^s\ +\ 
c\!\!\!\sum_{{\scriptsize \begin{array}{c}i\neq k\neq 
j\\(i>k)\end{array}}}\!\! (\bsigma_{i} \cdot \bsigma_{k})(P_i^s +P_k^s)P_j^s
\ +\ dP_1^sP_2^sP_3^s \nonumber
\end{eqnarray}
where the notation is defined in [2e]; $P_{i}^s$'s are the projectors 
on the strange quarks; $M_{0},\ B,\ C,\ \ldots,\ d$ are parameters. Of the two 
parameters $a$ and $b$ only the combination $(a+b)$ intervenes.

A comment on (\ref{4}): Because the different masses of the lowest 
octet and decuplet baryons are 8 (barring e.m. and isospin 
corrections), Eq.(\ref{4}), with 8 parameters 
$(M_{0},\ B,\ C,\ D,\ E,\ a+b,\ c,\ d)$, is certainly true, no matter 
what is the underlying theory. Yet the general parametrization (\ref{4}) is not 
trivial: The values of the above 8 parameters are seen to decrease 
strongly on moving to terms with increasing number of indices 
(Eq.(\ref{5})). In deriving (\ref{4}) from QCD, the term $\Delta m 
\bar{\psi}P^s \psi$ in the QCD Lagrangian is treated exactly; Eq. 
(\ref{4}) is correct to all orders in flavor breaking and the 
derivation takes into account all possible closed loops. In (\ref{4}) 
the parameters (in MeV) are -Ref.[3a]:
\begin{equation}
    \begin{array}{lclclcl}
 	\label{5}
 	M_{0}=1076 &,& B=192 &,& C=45.6 &,& D=-13.8\pm 0.3 \\
 	(a+b)=-16\pm 1.4 &,& E=5.1\pm 0.3 &,& c=-1.1\pm 0.7 &,& d=4\pm 3
 	\end{array}
 \end{equation} 
The hierarchy of these numbers is evident and, as shown in [3a], it 
corresponds \cite{N1} to a reduction factor $\approx 1/3$ for an 
additional pair of indices and $\approx 1/3$ for each flavor breaking 
factor $P_{i}^s$. The values (\ref{5}) decrease strongly with 
increasing complexity of the accompanying spin-flavor structure. 
Barring $c$ and $d$, the following mass formula results [2e], a 
generalization of the Gell-Mann Okubo formula that includes octet and 
decuplet:
\begin{equation}
\frac{1}{2} (p+\Xi^{0})+T=\frac{1}{4} (3\Lambda +2\Sigma^{+}-\Sigma^{0})
\label{6}
\end{equation}
The symbols stay for the masses and $T$ is the following 
combination of decuplet masses:
\begin{equation}
	T=\Xi^{\ast -} - \frac{1}{2} (\Omega + \Sigma^{\ast -})
\end{equation}
Because of the level of accuracy reached in comparing 
Eq.(\ref{6}) with the data, we wrote (\ref{6}) so as to be 
free of electromagnetic effects. (It can be easily checked the 
combinations in (\ref{6}) are independent of electromagnetic and 
isospin  effects, to zero order in flavor breaking.) The data satisfy 
(\ref{6}) as follows:
\begin{equation}
l.h.s.=1133.1\pm 1.0 \qquad r.h.s.=1133.3\pm 0.04
\end{equation}
an impressive agreement confirming the smallness of the 
terms neglected in (\ref{4}).

One more remark [3a]: A QCD 
calculation, if feasible, would express each ($M_{0},\ B \ldots c,\ d$) 
in (\ref{4}) in terms of the quantities in the QCD Lagrangian, the 
running quark masses -normalized at any $q$ that we like to select- and 
the dimensional (mass) parameter $\Lambda \equiv \Lambda_{QCD}$; for 
instance, setting for simplicity $m_{u}=m_{d}=m$:
\begin{equation}
M_{0}\equiv \Lambda \hat M_{0}(m(q)/\Lambda ,m_{s}(q)/\Lambda)
\end{equation}
where $\hat M_{0}$ is some function. Similarly for $B,\ C,\ D,\ E,\ a,\ 
b,\ c,\ d$ . The numerical value of the coefficients should be seen as 
the result of a QCD exact calculation performed with an arbitrary 
choice of the renormalization point of the running quark masses.
\vskip 40 pt
\noindent {\bf 3. A comparison with the large $N_{c}$ method.}
\vskip 5 pt

We now compare the parametrized baryon mass (\ref{4}), with that 
obtained in the $1/N_{c}$ method. There (compare ref.\cite{jl}, 
Eq.3.4) the parametrization of the baryon masses is also expressed in 
terms of 8 parameters (from $c_{(0)}^{1,0}$ to $c_{(3)}^{64,0}$), but, 
note, the quantities they multiply are collective rather than 
individual quark variables. It is again true that, setting to zero the 
smaller coefficients, one finds a relation (Eq.(4.6) in \cite{jl}) 
between octet and decuplet baryon masses, which is equivalent to Eq.
(\ref{6}).

Neither in Ref.\cite{jl} nor in other papers \cite{j} it was stated 
\cite{N2} that this relation coincides -except for the notation- with 
(\ref{6}), published long before. I note only, here, the following: The 
general QCD  parametrization can reproduce the good results of 
$1/N_{c}$ simply using, as we did, the conjecture that the empirical 
hierarchy apparent in Eqs. (\ref{4},\ref{5}) for the baryon masses, 
applies to many or all properties, at least for the lowest baryons 
with a factorizable $\phi_{B}$. It is, I repeat, what we always did 
in \cite{m,dm} (see [2g], fig.1). We explained \cite{m,dm} in this way 
a variety of facts about the magnetic moments, $\Delta^{+}\to 
p\gamma$ , semileptonic decays and many other quantities. Note that 
the GP method in principle includes all diagrams, not only the planar 
ones; the closed loops, related in the GP to the Trace terms (see in 
[3a], the ref.14) are also taken into account; their contribution is 
depressed or not depending on the number of additional gluons that 
are necessary, due to the Furry theorem (see [3c], in particular 
fig.1). For instance the Trace terms that were written in Ref.[3f] 
are depressed by the Furry theorem and can be neglected as we did. 
The phrasing of Buchmann and Lebed \cite{bl} did not clearly express 
this.

By the way, on reading Ref.\cite{bl} it looks as if by neglecting, as 
we did, from the start, terms proportional to $m_{u}-m_{d}$ 
(which are of the order 
$|m_{u}-m_{d}|/(3\Lambda_{QCD})\approx 5\cdot 10^{-3}$) we had imposed a 
``mild physical constraint''. This is not so. But, except for these 
points of language, Buchmann and Lebed seem to state that the 
relationship between the radii of $N$ and $\Delta$ implied by the GP 
and $1/N_{c}$ methods is the same. One may ask: Is it then really 
necessary to start from $N_{c}=\infty$?

Finally I comment on the Coleman-Glashow (CG) relation considered in 
a ref.[3h](see also \cite{N3} ). In ref.[3h], we recalled the GP result of 
ref.[2f] (only three index flavor breaking terms violate the CG 
relation) and showed that neither the $u-d$ mass difference, nor the 
Trace terms modify this conclusion. This explains the ``miracolous'' 
precision of the CG relation, which neglects entirely flavor breaking 
in its original derivation; such a precision is much better tested after a 
recent measurement of the $\Xi^{0}$ mass \cite{ex}. 

After the appearance of [3h](as hep-ph/004198,20 apr 2000)
a preprint by Jenkins and Lebed 
\cite{nat} implied by its title that in the large $N_{c}$ description 
it is quite natural (not "miracolous") that the CG relation is 
so beautifully verified. It is asserted in \cite{nat} that the 
neglected terms are naturally expected to be of an order in $1/N_{c}$ 
sufficiently high to guarantee their smallness.

This confidence, 
however, is not supported by their theory. E.g. the Trace terms 
present in the general parametrization (corresponding to closed loops) 
are many [3h]. Their negligible or vanishing global contribution 
cannot be established using only the order in $1/N_{c}$ of a typical term.
This is another reason,in addition  to the doubts raised in \cite{ss},
confirming  that it is not established that the
$1/N_{c}$ expansion can make predictions having a real QCD foundation.

{\em Acknowledgement.} I am very indebted to G.Dillon for frequent 
discussions.

\pagebreak


\begin{thebibliography}{99}
\bibitem{N1} An additional pair of indices in a term of the GP 
implies the exchange of an additional gluon; this carries a reduction 
factor $\approx 0.30$. This number is estimated using Eq.(\ref{4}) and 
taking the average between the values for the standard and pole 
masses (compare refs.[2e,3a]).
\end{thebibliography}
\end{document}